\documentclass[reprint,
superscriptaddress,
nofootinbib,
 amsmath,amssymb,
]{revtex4-2}

\usepackage{natbib,hyperref}

\usepackage{graphicx}
\usepackage{dcolumn}
\usepackage{bm}
\usepackage{hyperref}

\usepackage{autonum}
\usepackage{xcolor}
\usepackage{subcaption}

\renewcommand{\t}[1]{\tilde{#1}}

\begin{document}

\title{Ruling out unitary Fermi gas of neutrinos as galactic dark matter}

\author{Andr\'es Ace\~na}
\altaffiliation{acena.andres@conicet.gov.ar}
\affiliation{Instituto Interdisciplinario de Ciencias B\'asicas, CONICET, Facultad de Ciencias Exactas y Naturales, Universidad Nacional de Cuyo, Mendoza, Argentina.}

\begin{abstract}We investigate the possibility that galactic dark matter could be interacting fermions in the neutrino mass range. Assuming that galactic halos behave like a fluid in hydrostatic and thermal equilibrium, we employ the equation of state for a unitary Fermi gas to establish constraints on the mass of the fermion. Our findings effectively exclude neutrinos as candidates for galactic dark matter, and further reinforce earlier results suggesting that galactic dark matter is predominantly composed of light bosons.
\end{abstract}

\maketitle

\section{Introduction}

One of the outstanding problems in astrophysics, as well as in particle physics, is to elucidate the nature of dark matter (DM). The current cosmological model, $\Lambda$-CDM, indicates that DM is the dominant matter component in the universe \cite{Planck:2018vyg,Fields:2006ga,WMAP:2010qai}. It also plays a fundamental role in explaining galaxy structure, kinematics, formation, and evolution \cite{Springel:2005mi,Salucci:2018hqu}. So far, the most promising candidates to be DM particles are WIMPs \cite{Lee:1977ua, Bertone:2004pz}, although also sterile neutrinos \cite{Shi:1998km}, axions and light bosons \cite{Abbott:1982af}, and macroscopic objects like MaCHOs \cite{EROS-2:2006ryy} have been proposed. Nonetheless, experiments designed to detect the passage of DM particles through normal matter have provided negative results \cite{PandaX-4T:2021bab,LZ:2022lsv,XENON:2023cxc}, the same with indirect searches \cite{Barman:2022jdg,Chan:2019ptd,Song:2023xdk}, and results from the Large Hadron Collider exclude most models that offer plausible candidates \cite{CMS:2022qva,ATLAS:2022ygn}.

In the cosmological setting, DM is assumed to be a pressureless fluid. There are strong indications that it can not be baryonic matter \cite{Fields:2006ga,WMAP:2010qai}. At galactic scales, the pressureless hypothesis is unable to provide a complete description of the DM halos \citep{Weinberg:2013aya,Persic:1995ru,Salucci:2018hqu,Perivolaropoulos:2021jda}. Therefore, the DM problem is still open and several models beyond $\Lambda$-CDM have been proposed. However, most of them start with a theoretical hypothesis and draw upon on observations to determine the free parameters of the model.
As a way to narrow the parameter space in the search for the DM particle, a more observationally based type of investigation may be needed \cite{Nesti:2023tid}. For example, galactic rotation curves are an excellent laboratory where properties of DM can be extracted, as there is enough high precision data, showing several scaling relations that span over a huge range of sizes \cite{Kormendy:2004se,Donato:2009ab, Gentile:2009bw, Kormendy:2014ova}.

Considering said rotation curves and from a phenomenological standpoint, in \cite{Barranco:2013wy} a new model for DM was presented in the form of an equation of state (EOS), relating the pressure to the density. It was assumed that the DM halo could be modeled as a perfect fluid in hydrostatic equilibrium. The main issue with the EOS presented is that each galaxy corresponds to a different EOS, all with the same functional form but with different parameters. Crucially, in \cite{Acena:2021wjx} it was shown that the self-gravitating configurations corresponding to DM halos of actual galaxies constructed from those EOS are unstable. Following the same approach, in \cite{Acena2023} the temperature profiles of the DM halos were obtained, modeling DM as either an ideal gas, a Fermi or Bose gas. It was found that only for the Bose case the temperature profile decreases outwards, as generally expected from the thermodynamics point of view, and only if the mass of the boson is in the approximate range
\begin{equation}\label{mDM}
    10^{-3}\, \mbox{eV}/c^2 \lesssim m_{DM} \lesssim 10 \,\mbox{eV}/c^2.
\end{equation}
No known particle satisfies these requirements, but remarkably, neutrinos are expected to be in such a range.

The latest results from the KATRIN experiment \cite{Katrin2024} show that the electron-neutrino mass is bounded by
\begin{equation}
    m_\nu < 0.45\,\mbox{eV}/c^2.
\end{equation}
In the context of $\Lambda$-CDM, from the combination of different observational data sets, the bound
\begin{equation}
    \sum m_\nu <  0.072\,\mbox{eV}/c^2
\end{equation}
is obtained for the sum of neutrino masses \cite{Planck:2018vyg,Desi2024}. Furthermore, neutrino oscillation experiments have shown that at least two of the three
neutrino masses are non-zero, but the ordering of these masses is not known \cite{Gonzalez2021}. In the normal hierarchy
\begin{equation}
    \sum m_\nu < 0.059\,\mbox{eV}/c^2,
\end{equation}
while in the inverted hierarchy
\begin{equation}
    \sum m_\nu < 0.10\,\mbox{eV}/c^2.
\end{equation}

Connecting the results in \cite{Acena2023} with the bounds on neutrino masses, it is tempting to postulate that DM is composed of neutrinos, but they are fermions, which were discarded in that work. Noteworthy, gases composed of interacting fermions can have bosonic behavior in certain situations. Also, from the analysis in \cite{Acena2023}, it is clear that quantum effects in the DM halo can be of fundamental importance. Therefore, in the present work, we explore the possibility that galactic halos could be composed of interacting neutrinos. This entails trying to reproduce the physical parameters of the DM halos with a fermionic gas whose particles are in the approximate mass range \eqref{mDM}.

Our hypothesis regarding galactic DM are the same as in \cite{Barranco:2013wy,Acena:2021wjx,Acena2023}. We consider that the DM halos can be modeled as a fluid that forms a self-gravitating structure in hydrostatic equilibrium, which then uses the baryonic matter as tracer of its gravitational potential. This last assumption is particularly accurate for Low Surface Brightness (LSB) galaxies, which appear to be completely dominated by DM, even in the inner regions \cite{McGaugh1998}. Firstly, in an exploratory fashion, we analyze the possibility that there is an unknown interaction among neutrinos that gives rise to tightly bound dimers. The existence of neutrino-bound states and multi-particle systems has been extensively discussed in \cite{Smirnov2022}. Although we conclude that the bounds on neutrino-neutrino interactions exclude the possibility that galactic DM could be composed of neutrino dimers, the analysis clarifies the situation in which neutrinos would be if they were galactic DM. Fundamentally, the thermal wavelength would be bigger than the dimer radius, which itself would be bigger than the distance between dimers. This situation is indicative of the gas being in the crossover between the Bardeen-Cooper-
Schrieffer (BCS) and the Bose-Einstein Condensate (BEC) phases \cite{Garani2022}. In this crossover there is a particular situation of special importance, the unitary Fermi gas (UFG).

The UFG describes a system of fermions (such as neutrons or atoms) interacting through a zero-range potential, where the scattering length is infinite. This results in strong interactions between the particles.  At unitarity, the system exhibits universal behavior, meaning that many properties are independent of the details of the interactions. This includes aspects like thermodynamic properties and the EOS.  The UFG can exhibit various phases, such as superfluidity, depending on temperature and density.
Therefore, the thermodynamic properties of the UFG can be obtained using any one realization of it. In particular, they can be determined experimentally using ultracold diluted gases of fermionic alkali atoms tuned via a Feshbach resonance \cite{Zwerger2011}.  Important for us is that galactic DM being a UFG of neutrinos is compatible with neutrinos interacting only through weak interactions. This is particularly appealing, as DM could be explained by resorting only to currently known particles and interactions. Furthermore, the UFG has properties that could explain some of the special properties that DM has. For example, the bulk viscosity of the UFG vanishes identically and in the superfluid regime, there is a steep drop in the shear viscosity \cite{Hou2021}. Nevertheless, our analysis shows that the UFG of neutrinos has to be discarded. A particle mass well above current upper bounds on neutrino masses is needed to be able to reproduce the DM halo profiles.

Our main conclusion is that neutrinos can not be galactic DM. Furthermore, fermions in the mass range \eqref{mDM} can only be DM if the interaction is so strong that the dimer is effectively a bosonic particle. This in fact strengthens the conclusions obtained in \cite{Acena2023}, where the DM particle is proposed to be a boson in said mass range.

\section{The PSS profile, hydrostatic equilibrium and thermodynamic considerations}

In the same line as \cite{Barranco:2013wy,Acena:2021wjx,Acena2023}, we start by considering the rotational velocity profile proposed by Persic, Salucci, and Stel \cite{Persic:1995ru}, called the PSS profile or the Universal Velocity profile. It has two parameters, $v_\star$ and $r_\star$, where $v_\star$ is the flat terminal velocity and $r_\star$ is a measure of the extent of the central region in which the velocity has a solid body profile $v\propto r$. Although several velocity or density profiles for DM halos have been proposed, the PSS profile has the advantage that it faithfully reproduces both the central region and the flat velocity region, which extends far away from the luminous matter \cite{Mistele2024}.

\begin{figure}[h]
    \centering
    \includegraphics[width=\linewidth]{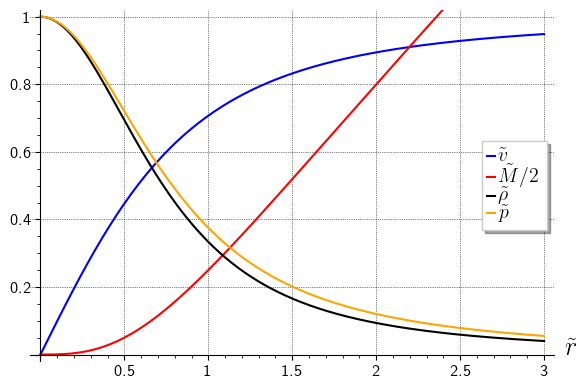}
    \caption{Universal profiles corresponding to the PSS rotational velocity profile.}
    \label{figperfiles}
\end{figure}

The PSS profile is given by the analytical expression
\begin{equation}
    v = \frac{v_\star r}{\sqrt{r^2+r_\star^2}}.
\end{equation}
Having $v(r)$, the mass distribution is obtained from $M = \frac{rv^2}{G}$, and then the density profile is $\rho = \frac{1}{4\pi r^2}\frac{dM}{dr}$. Finally, the hydrostatic equilibrium equation, $\frac{dp}{dr}=-\frac{GM\rho}{r^2}$, is solved to obtain the pressure profile. Furthermore, we consider the relation obtained in \cite{Kormendy:2004se}, later confirmed and extended in \cite{Donato:2009ab, Gentile:2009bw, Kormendy:2014ova}, which in terms of the PSS profile is\footnote{The value presented is $\rho_0 r_0 = 141^{+82}_{-52}\,M_\odot/\mbox{pc}^2$, being $r_0$ the parameter corresponding to the Burkert density profile, which is related to the PSS profile through $r_0\approx 1.5 \,r_\star$.}
\begin{equation}\label{kormendy}
 \rho_0 r_\star = \eta \approx 94\, M_\odot/\mbox{pc}^2.
\end{equation}
This allows us to reduce one parameter in the PSS profile. We will keep $v_\star$ and then
\begin{equation}
    r_\star = \frac{3v_\star^2}{4\pi G\eta} \approx 5.9\,\mbox{kpc} \times \left(\frac{v_\star}{100\,\mbox{km}/\mbox{s}}\right)^2.
\end{equation}
We define the representative quantities
\begin{align}
& M_\star = \frac{3v_\star^4}{4\pi G^2\eta} \approx 1.4\times 10^{10}\,M_\odot \times \left(\frac{v_\star}{100\,\mbox{km}/\mbox{s}}\right)^4, \\
& \rho_\star = \frac{4\pi G\eta^2}{3 v_\star^2} \approx 0.60 \,\frac{\mbox{GeV}/c^2}{\mbox{cm}^3}\times \left(\frac{v_\star}{100\,\mbox{km}/\mbox{s}}\right)^{-2}, \\
& p_\star = \frac{4}{9}\pi G \eta^2 \approx 2.0\times 10^9 \,\frac{\mbox{GeV}}{\mbox{cm}^3},
\end{align}
and the dimensionless variables
\begin{equation}
    \t{r} = \frac{r}{r_\star},\quad \t{M} = \frac{M}{M_\star}, \quad \t{\rho} = \frac{\rho}{\rho_\star},\quad \t{p} = \frac{p}{p_\star}.
\end{equation}
The profiles in these variables are universal,
\begin{align}
    & \t{v} = \frac{\t{r}}{\sqrt{\t{r}^2+1}}, \quad \t{M} = \frac{\t{r}^3}{\t{r}^2+1}, \\
    & \t{\rho} = \frac{\t{r}^2+3}{3(\t{r}^2+1)^2}, \quad \t{p} = \frac{\t{r}^2+2}{2(\t{r}^2+1)^2},
\end{align}
and they are shown in Fig. \ref{figperfiles}. It is interesting to note that in our context, the relationship \eqref{kormendy} implies that $p(r=0)=p_\star$ is the same for all galactic halos, which was first noticed in \cite{Acena:2021wjx}.

The equilibrium thermodynamics implications of these profiles were studied in \cite{Acena2023}, where the matter model used corresponded to a classical ideal gas, an ideal Fermi gas and an ideal Bose gas. It was shown that the only model consistent with a decreasing temperature profile was the Bose one, and that at least the central region of the halo is in a BEC. For the condensed region the temperature profile is
\begin{align}
    T  & = \frac{1}{k}\left(\frac{h^6p^2}{8\pi^3 \zeta^2 \left(\frac{5}{2}\right) m_{DM}^3}\right)^\frac{1}{5} \\
    & = \frac{1}{k}\left(\frac{2 h^6 G^2 \eta^4}{3^4 \pi \zeta^2 \left(\frac{5}{2}\right) m_{DM}^3}\frac{(\t{r}^2+2)^2}{4(\t{r}^2+1)^4}\right)^\frac{1}{5}.
\end{align}
As noted in \cite{Acena2023}, the relation \eqref{kormendy} implies that the central temperature of all galactic DM halos is the same, and for the condensed region the profile is also universal. We define the representative central temperature
\begin{equation}
    T_* = \frac{1}{k}\left(\frac{2 h^6 G^2 \eta^4}{3^4 \pi \zeta^2 \left(\frac{5}{2}\right) m_{DM}^3}\right)^\frac{1}{5} \approx 0.24\,\mbox{K} \times \left(\frac{m_{DM}}{\mbox{eV}/c^2}\right)^{-\frac{3}{5}}.
\end{equation}
For comparison, if we consider the halo to be composed of a classical ideal gas, then the temperature profile is increasing, and asymptotically it is isothermal, with temperature
\begin{equation}
    T_{iso} = \frac{m_{DM} v_\star^2}{2k} \approx  6.5\times 10^{-4}\,\mbox{K} \times \left(\frac{m_{DM}}{\mbox{eV}/c^2}\right) \left(\frac{v_\star}{100\,\mbox{km}/\mbox{s}}\right)^2.
\end{equation}
Crucially, in \cite{Acena2023} it is concluded that for the halo to be well-behaved thermodynamically, the DM particle has to be a boson with mass in the range \eqref{mDM}.

In the next section we use that from these considerations we have the density, pressure and temperature, and a permissible range for the mass of the particle, to test the hypothesis that DM is composed of strongly interacting neutrinos.

\section{Neutrinos with Yukawa coupling}

Given the results in \cite{Acena2023}, the easiest way for neutrinos to be the DM particle would be if there were an interaction strong enough to form neutrino dimers. Since we can obtain at most estimates of orders of magnitude, we consider only one neutrino species, with mass $m_\nu$, and therefore the dimer would have mass $m_{DM} = 2 m_\nu$. For a discussion of neutrino-bound states, see \cite{Smirnov2022}. If there is an attractive interaction among neutrinos, mediated by a particle of mass $m_\phi$, with $m_\phi$ sufficiently smaller than $m_\nu$, then the Yukawa potential is
\begin{equation}
    V = -\frac{\hbar c \alpha}{r}exp\left(-\frac{m_\phi}{\hbar c} r\right),
\end{equation}
where $\alpha$ is the dimensionless coupling constant. The condition for the existence of at least one bound state is
\begin{equation}
    \frac{\alpha m_\nu}{2 m_\phi} > 0.8499.
\end{equation}
The most favorable situation for the formation of dimers is the limit $m_\phi=0$, in which case
\begin{equation}
    V = -\frac{\hbar c \alpha}{r}.
\end{equation}
The corresponding binding energy of the ground state is
\begin{equation}
    E = -\frac{1}{4} \alpha^2 m_\nu c^2,
\end{equation}
with radius
\begin{equation}
    a = \frac{2\hbar}{\alpha m_\nu c}.
\end{equation}
As a comparison, we will use the nearest neighbor radius, $r_{NN}$, which is a measure of closeness for particles in the gas,
\begin{equation}
    r_{NN} = \left(\frac{m_{DM}}{2\pi \rho}\right)^\frac{1}{3}.
\end{equation}
Also, we consider the thermal wavelength of the DM particle,
\begin{equation}
    \lambda = \frac{h}{\sqrt{2\pi m_{DM} k T}}.
\end{equation}

\begin{figure}[h]
    \centering
    \includegraphics[width=\linewidth]{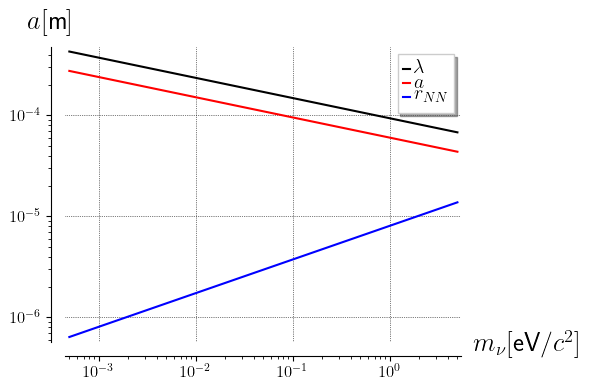}
    \caption{Characteristic distances among particles for $v_\star = 150\,\mbox{km}/\mbox{s}$.}
    \label{figdistances}
\end{figure}

To estimate a reasonable value for $\alpha$ we take the criteria that the average kinetic energy per particle not in the condensed state has to be lower than the ground state energy, that is, $K<|E|$, being
\begin{equation}
    K = \frac{3}{2}kT\frac{\zeta\left(\frac{5}{2}\right)}{\zeta\left(\frac{3}{2}\right)},
\end{equation}
and therefore
\begin{align}
    & \alpha > \alpha_K, \\
    & \alpha_K = \left(\frac{6kT_\star}{c^2 m_\nu}\frac{\zeta\left(\frac{5}{2}\right)}{\zeta\left(\frac{3}{2}\right)}\right)^\frac{1}{2} \approx 6.5 \times 10^{-3}  \times \left(\frac{m_\nu}{\mbox{eV}/c^2}\right)^{-\frac{4}{5}}.
\end{align}
We consider the range
\begin{equation}
    5\times 10^{-4}\, \mbox{eV}/c^2 \lesssim m_\nu \lesssim 5 \,\mbox{eV}/c^2,
\end{equation}
and then we see that $\alpha_K$ is much bigger than the allowed experimental bounds collected in \cite{Smirnov2022}. Since the bound on $\alpha$ is most restrictive at the origin, we evaluate $a$, $r_{NN}$ and $\lambda$ there, obtaining
\begin{align}
    a & = \left(\frac{2\hbar^4 \zeta^5\left(\frac{3}{2}\right)}{3\pi^5 G^2 \eta^4 \zeta^3\left(\frac{5}{2}\right) m_\nu^2}\right)^\frac{1}{10} \\
    & \approx 6.0 \times 10^{-5}\,\mbox{m}  \times \left(\frac{m_\nu}{\mbox{eV}/c^2}\right)^{-\frac{1}{5}}, \\
    r_{NN} & = \left(\frac{3 m_\nu v_\star^2}{4\pi^2 G\eta^2}\right)^\frac{1}{3} \\
    & \approx 8.1 \times 10^{-6}\,\mbox{m}  \times \left(\frac{m_\nu}{\mbox{eV}/c^2}\right)^\frac{1}{3} \left(\frac{v_\star}{100\,\mbox{km}/\mbox{s}}\right)^\frac{2}{3}, \\
    \lambda & = \left(\frac{3^4 \hbar^4 \zeta^2\left(\frac{5}{2}\right)}{2^4 G^2 \eta^4 m_\nu^2}\right)^\frac{1}{10} \\
    & \approx 9.4 \times 10^{-5}\,\mbox{m}  \times \left(\frac{m_\nu}{\mbox{eV}/c^2}\right)^{-\frac{1}{5}}.
\end{align}
In Fig. \ref{figdistances} we present $a$, $r_{NN}$ and $\lambda$ as functions of $m_\nu$ for a mid-size halo with $v_\star = 150\,\mbox{km}/\mbox{s}$. We see that for the ranges of parameters considered,
\begin{equation}
    r_{NN}<a<\lambda.
\end{equation}
This is the opposite of what would be expected for a gas of strongly bonded dimers. Together with the experimental bounds on $\alpha$, this renders the modeling of galactic DM as neutrino dimers unfeasible. At the same time, it hints at another possibility. The only known interaction of neutrinos among themselves is the weak interaction. The range of such interaction is too short to form bound states, but at the same time the thermal wavelength for neutrinos as DM would be bigger than the neutrino-neutrino spacing. This points to the possibility of having an UFG, which we explore in the next section.

\section{UFG}

From the previous considerations, if we assume that galactic DM consists of neutrinos, we are in a situation where
\begin{equation}
    r_I << r_{NN} < \lambda,
\end{equation}
being $r_I$ the interaction range for the weak force. This does not automatically imply that we are dealing with a UFG, but hints in this direction.

\begin{figure}[h]
    \centering
    \includegraphics[width=\linewidth]{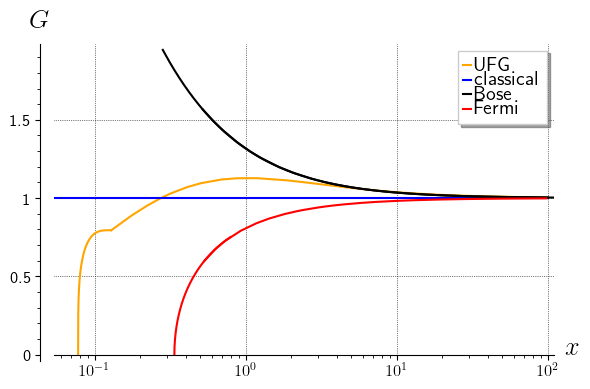}
    \caption{Correction factor for the EOS.}
    \label{figfactorG}
\end{figure}

The EOS for the UFG was presented in \cite{Hou2021}. It can be written similarly to the EOS for the Fermi gas in the form
\begin{equation}
 T = G(x)\frac{m}{k}\frac{p}{\rho},\quad x = \frac{2m^4}{h^3}\left(8\pi^3\frac{p^3}{\rho^5}\right)^\frac{1}{2},
\end{equation}
where $m$ is the mass of the fermion.
There is an absolute minimum value for $x$,
\begin{equation}
 x_0 = 0.0775,
\end{equation}
and a critical value, $x_c$, where the transition to a superfluid phase occurs,
\begin{equation}
 x_c = 0.1285.
\end{equation}
The function $G(x)$ is defined in the two regions:
\begin{equation}
 G(x) = \frac{1+d_1/x+d_2/x^2}{1+d_3/x+d_4/x^2},\quad x>x_c,
\end{equation}
where
\begin{equation}
 d_1 = 1.8052,\,d_2 = -0.0022,\,d_3 = 1.3668,\,d_4 = 0.1179,
\end{equation}
and
\begin{equation}
 G(x) = G_c\left[1-\left(\frac{x_c-x}{x_c-x_0}\right)^4\right]^\frac{1}{4}, \quad x_0\leq x \leq x_c,
\end{equation}
\begin{equation}
 G_c = 0.7944.
\end{equation}
From this we see that for a given $p$ there is maximum $\rho$,
\begin{equation}\label{condRho}
 \rho < \rho_{max,0}, \quad \rho_{max,0} = \left(\frac{2^5\pi^3 m^8 p^3}{h^6 x_0^2}\right)^\frac{1}{5}.
\end{equation}
Also, for being in the normal phase
\begin{equation}
 \rho < \rho_{max,c}, \quad \rho_{max,c} = \left(\frac{2^5\pi^3 m^8 p^3}{h^6 x_c^2}\right)^\frac{1}{5},
\end{equation}
and the associated temperature is
\begin{equation}
 T_c = \frac{G_c}{2k}\left(\frac{h^6 x_c^2 p^2}{\pi^3 m^3}\right)^\frac{1}{5}.
\end{equation}
In Fig. \ref{figfactorG} we present the factor $G$ as a function of $x$, together with the same factor for a classical ideal gas, for an ideal Fermi gas and for an ideal Bose gas. We see that regarding the EOS, for $x\gtrsim 2$ the behavior of the UFG resembles more a Bose than a Fermi gas. For $x\lesssim 0.5$ the behavior resembles a Fermi gas, although displaced to lower values of $x$. This gives some hope that an UFG EOS would be able to better reproduce the DM halo than an ideal Fermi gas does.

Regarding the PSS profile, this EOS does not give an outwards increasing temperature profile, and in this sense it has to be discarded on the same grounds as the classical ideal gas and the ideal Fermi gas. But there are still stronger reasons. If we have the density and pressure profiles, then we need that at least
\begin{equation}
    m > \left(\frac{h^6 x_0^2 \rho^5}{2^5 \pi^3 p^3}\right)^\frac{1}{8}.
\end{equation}
For the PSS profile this means that
\begin{equation}
 m > \left(\frac{3 h^6 G^2 \eta^4 x_0^2}{2 \pi v_\star^{10}}\right)^\frac{1}{8} \approx 25 \,\mbox{eV}/c^2 \times \left(\frac{v_\star}{100\,\mbox{km}/\mbox{s}}\right)^{-\frac{5}{4}}.
\end{equation}
If we take $x_0\rightarrow x_c$ then the condition on $m$ is for the non-appearance of a superfluid phase. The lower bound on $m$ leaves out the possibility of reproducing the PSS profile with neutrinos, even if they form a UFG.

\begin{figure}[h]
    \centering
    \includegraphics[width=\linewidth]{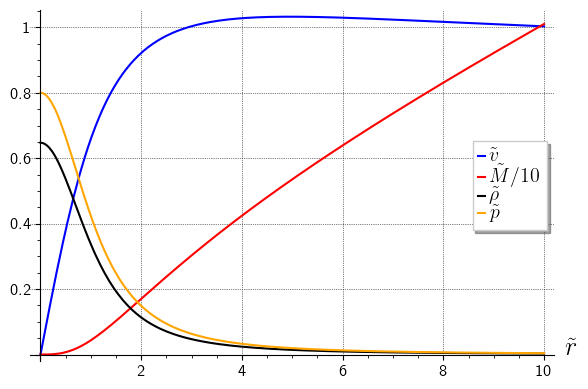}
    \caption{Profiles for an UFG isothermal halo with $m = 25\,\mbox{eV}/c^2$, $\t{\rho}_0 = 0.65$ and $\t{p}_0=0.8$.}
    \label{figperfilesIsoNu}
\end{figure}

Being that the PSS profile with the neutrino UFG as matter model can not be reproduced, we consider the possibility of an isothermal profile. Again, we have the relationship \eqref{condRho}, which can be rewritten as
\begin{align}
     \t{p}_0 & > \left(\frac{3 h^6 G^2 \eta^4 x_0^2}{2\pi m^8 v_\star^10} \t{\rho}_0^5\right)^\frac{1}{3} \\
    & \gtrsim 5.4 \times 10^{3} \times \left(\frac{m}{\mbox{eV}/c^2}\right)^{-\frac{8}{3}} \left(\frac{v_\star}{100\,\mbox{km}/\mbox{s}}\right)^{-\frac{10}{3}} \t{\rho}_0^\frac{5}{3}.
\end{align}
This implies that for $m$ in the range \eqref{mDM} the values of $\t{\rho}_0$ and $\t{p}_0$ are far away from the values needed to reproduce a reasonable halo profile, which confirms that neutrinos can not be galactic DM. For $m$ above certain values, the profiles can be reproduced, but the bounds on $m$ exclude neutrinos. As an example, in Fig. \ref{figperfilesIsoNu} we present the profiles for the values $m = 25\,\mbox{eV}/c^2$, $\t{\rho}_0 = 0.65$ and $\t{p}_0=0.8$, which gives an isothermal halo with temperature $T = 0.034\,\mbox{K}$.

\section{Conclusion}

Our main conclusion is that neutrinos can not be galactic DM. If we take the results in \cite{Acena2023}, together with the present analysis, the conclusion is that the DM particle is either a boson or a strongly interacting fermion, for which no particle candidate has been detected. In the boson case, the mass window corresponds to the upper end of the QCD axion \cite{Chadha-Day2022,Semertzidis2022}. Although we have used the PSS profile, using other profiles proposed in the literature does not fundamentally change the obtained parameters for the DM halos, and the conclusion still holds. Given that an ideal Bose gas does give a thermodynamically satisfactory description of the DM halos, while the ideal Fermi gas does not, and that the UFG, while also not satisfactory, is an improvement over the ideal Fermi gas, one can think of using an EOS for an interacting Fermi gas closer to the BEC side in the BCS-BEC crossover. This means that one could manufacture an EOS with the desired properties but, given the experimental upper bounds on the neutrino mass, this also means that an extremely delicate fine-tuning is necessary in the low temperature/high density regime, and this seems impossible to justify on physical grounds.

We again find that the key difference between bosons and fermions is essential in the study of DM. Bosons, which adhere to Bose-Einstein statistics, can occupy the same quantum state, enabling the formation of condensates that exhibit long-range interactions. This feature makes bosons particularly attractive as DM candidates, especially in the context of scalar fields or axions, which may possess the necessary properties for a stable DM component.
On the other hand, neutrinos, being fermions, follow the Pauli exclusion principle. This principle prevents two fermions from occupying the same quantum state at the same time. As a result, neutrinos cannot effectively cluster under gravitational forces to form the large-scale structures observed in the universe. Furthermore, their weak interactions with matter and relatively low mass complicate their viability as DM candidates.
These insights suggest that theories centered on bosonic DM models are more promising. Such models could potentially account for phenomena like self-interacting DM or coherent DM waves, which may help explain various cosmic observations, including galaxy distributions and the dynamics of galaxy clusters.
By emphasizing the limitations of neutrinos as candidates for DM, we bolster our case for bosons, encouraging further investigation into their properties and interactions. This underscores the potential of exploring bosonic fields to provide valuable insights into the nature of DM and its significance in the universe.

In relation to the DM halos, our results do not imply that other hypotheses regarding their physical situation or the properties of the DM particle can not produce reasonable halos. For example, a polytropic EOS produces halo profiles that fit the observationally obtained rotation curves \cite{Novotny2021}. In this regard, we consider that the strength of our approach is that with a few physically motivated assumptions, a good deal of microscopic information can be extracted from the astrophysical data of galaxies.

\vspace{0.5cm}

\section*{Acknowledgments}

I thank Juan Barranco, Argelia Bernal, and Ericson Lopez for fruitful discussions and comments on the manuscript.
This work was partially supported by Agencia I+D+i (Argentina) through grant PICT-2021-I-INVI-00597.
Part of the calculations, including the numerical integrations needed to obtain the profiles, were performed using SageMath \cite{sagemath}.





\bibliography{biblio}
\bibliographystyle{abbrvnat}

\end{document}